\documentclass{amsart}
\usepackage{amsfonts,amssymb,amsmath,epstopdf}
\usepackage{hyperref}
\usepackage{cleveref}
\usepackage{color,bm} 
\usepackage{tikz}
\hypersetup{backref,hyperindex=true,pageanchor=true,colorlinks=true,citecolor=blue,urlcolor=blue}

\newcommand{\R}{\mathbb{R}}

\newcommand{\wt}{\tilde{\omega}}
\newcommand{\kt}{\tilde{k}}
\newcommand{\qt}{\tilde{q}}
\newcommand{\phit}{\tilde{\varphi}}

\newcommand{\uv}{\mathbf{u}}

\newcommand{\x}{\mathbf{x}}

\newcommand{\ud}{\, \mathrm{d}}
\newcommand{\oth}{\overline{\theta}}
\newcommand{\omu}{\overline{\mu}}
\newcommand{\orho}{\overline{\rho}}
\newcommand{\ou}{\overline{u}}
\newcommand{\ov}{\overline{v}}

\DeclareRobustCommand\mysquare{\tikz \fill[black] (0,0) rectangle (0.14cm,0.14cm);}
\DeclareRobustCommand\mycircle{\tikz \fill[black] (0,0) circle (0.07cm);}
\DeclareRobustCommand\mytriangle{\tikz \fill[black] (0,0)--(0.18cm,0)--(0.09cm,0.156cm)--cycle;}

\DeclareRobustCommand\mypentagon{\tikz \fill[black]
  (0:0.08cm) \foreach \x in {18,90,...,306} { -- (\x:0.08cm) } -- cycle;}

\newcommand{\TheTitle}{Oblique spatial dispersive shock waves in
  nonlinear Schr\"{o}dinger flows}

\title{{\TheTitle}}
\thanks{M.A.H. was partially supported by NSF CAREER DMS-1255422 and
  DMS-1008973.  A.M.K.~was partially supported by RFBR grant No.~16-01-00398.}

\author{ M.~A.~Hoefer}
\thanks{Department of Applied Mathematics,
    University of Colorado Boulder, Boulder, CO
    (\email{hoefer@colorado.edu}} 
\author{G.~A.~El}
\thanks{Department of Mathematical Sciences, Loughborough
    University, Loughborough LE11 3TU, United Kingdom
    (\email{g.el@lboro.ac.uk})} 
\author{A.~M.~Kamchatnov}
  \thanks{Institute of Spectroscopy, Russian Academy of Sciences,
    Troitsk, Moscow Region, 142190, Russia
    (\email{kamch@isan.troitsk.ru})}

\begin{document}

\maketitle

\begin{abstract}
  In dispersive media, hydrodynamic singularities are resolved by
  coherent wavetrains known as dispersive shock waves (DSWs).  Only
  dynamically expanding, temporal DSWs are possible in one-dimensional
  media.  The additional degree of freedom inherent in two-dimensional
  media allows for the generation of time-independent DSWs that
  exhibit spatial expansion.  Spatial oblique DSWs, dispersive analogs
  of oblique shocks in classical media, are constructed utilizing
  Whitham modulation theory for a class of nonlinear Schr\"{o}dinger
  boundary value problems.  Self-similar, simple wave solutions of the
  modulation equations yield relations between the DSW's orientation
  and the upstream/downstream flow fields.  Time dependent numerical
  simulations demonstrate a convective or absolute instability of
  oblique DSWs in supersonic flow over obstacles.  The convective
  instability results in an effective stabilization of the DSW.
\end{abstract}



\section{Introduction}
\label{sec:introduction}

Breaking of hydrodynamic flows in conservative, weakly dispersive
media yields dispersive shock waves (DSWs) characterized by expanding
nonlinear wavetrains.  Mathematically, a great deal is known about
dispersive shock waves in the framework of nonlinear wave modulation
theory, also known as Whitham theory
\cite{whitham_non-linear_1965,gurevich_nonstationary_1974}, and
inverse scattering
\cite{lax_small_1983,venakides_long_1986,ablowitz_interactions_2013}
for spatio-temporal (1+1)-dimensional systems (cf.~the review
\cite{el_dispersive_2016}).  The limited theoretical investigations of
DSWs in multiple dimensions
\cite{gurevich_supersonic_1995,gurevich_supersonic_1996,el_spatial_2006,klein_numerical_2007,el_two-dimensional_2009,hoefer_theory_2009,hoefer_dark_2012,kartashov_two-dimensional_2013,klein_numerical_2013,ablowitz_dispersive_2016,dubrovin_critical_2016}
and the experimental realization of multi-dimensional DSWs in
ultracold atoms as superfluid matter waves
\cite{dutton_observation_2001,simula_observations_2005,hoefer_dispersive_2006}
and in nonlinear optical diffraction patterns
\cite{wan_dispersive_2007,ghofraniha_shocks_2007} provides motivation
for this study of two-dimensional DSWs.  Furthermore, with the
exception of recent studies on Kadomtsev-Petviashvili and related
(2+1)D equations
\cite{klein_numerical_2007,klein_numerical_2013,ablowitz_dispersive_2016,dubrovin_critical_2016},
the remaining previous theoretical works invoke asymptotic reductions
to the Korteweg-de Vries or (1+1)D nonlinear Schr\"{o}dinger (NLS)
equations.  These equations' complete integrability enables a detailed
analytical description via the existence of Riemann invariants for the
associated Whitham modulation equations.  Whitham theory is applicable
to a much wider class of equations and a description of salient DSW
features is possible by making a simple wave assumption
\cite{el_resolution_2005}.  In this article, we use simple wave DSW
theory (DSW fitting \cite{el_dispersive_2016}) to construct large
amplitude, spatial oblique DSWs for the two-dimensional,
time-independent (2+0)D NLS equation.  A spatial oblique DSW inhabits
a wedge region in the plane filled with stationary, modulated periodic
waves.  The bounding angles of the oblique DSW are completely
determined by the upstream Mach number and the downstream flow angle.
Consequently, this asymptotic solution can be applied to the problem
of supersonic flow over a corner.  We investigate the dynamical
stability of an oblique DSW for the corner problem, numerically
observing instability.  Utilizing the instability theory of oblique
dark NLS solitons
\cite{kamchatnov_stabilization_2008,hoefer_dark_2012,kamchatnov_condition_2011},
we classify the convective or absolute nature of the instability.  The
former provides an effective dynamical stabilization of the oblique
DSW near the apex of the corner.  Because the 2D NLS equation is an
accurate model of nonlinear optical and matter wave dynamics, among
other systems, these results exhibit wide application.

\subsection{DSW Fitting}
\label{sec:simple-dsws}

Generally, Whitham averaging for an $n^{\textrm{th}}$ order (1+1)D
dispersive wave equation proceeds in several steps.  The first
requirement is an $n$-parameter family of periodic traveling wave
solutions.  The period and frequency of oscillation determines a
length and time scale associated with dispersively initiated dynamic
processes.  Whitham theory then proceeds by assuming the existence of
larger modulation length and time scales on which the traveling wave's
parameters adiabatically vary.  The first order, quasi-linear
modulation equations for these parameters can be determined by
averaging $n-1$ conservation laws associated to the dispersive wave
equation.  The system is closed by an additional modulation equation,
known as the conservation of waves, that results from a consistency
condition inherent to the assumption of adiabatic parametric
evolution.  An important property required for modulational stability
of the traveling wave is hyperbolicity of the modulation equations.
If the Whitham modulation equations exhibit singularity formation,
modulations of quasi-periodic or multiphase solutions may be used, if
they exist.  In this work, we apply this same procedure to a
$5^{\textrm{th}}$ order dispersive (2+0)D spatial system.  One of the
spatial variables can be viewed as a time-like variable hence, all of
the Whitham machinery applies.

Regularization of gradient catastrophe in dispersive hydrodynamics can
be asymptotically described utilizing Whitham theory.  Away from DSWs,
the solution is slowly varying and can be approximately described by
the dispersionless equations (dispersionless zone).  Shock formation
gives rise to the development of oscillations characterized by a
modulated traveling wave (oscillation zone).  The interfaces between
the oscillation and dispersionless zones are unknowns that must be
determined along with the solution.  Within the oscillation
(dispersionless) zone, the Whitham (dispersionless) equations are
solved.  The two zones are matched at their free boundaries by
equating the averaged solution to the dispersionless solution and
requiring that either the oscillation amplitude goes to zero (harmonic
limit) or the oscillation period goes to infinity (soliton limit).
Admissibility criteria analogous to entropy conditions for classical
shock waves determine the appropriate limiting behavior.  Thus a
complete asymptotic description of dispersive hydrodynamics involves
the determination of the space-time dynamics of all the modulation
parameters and the boundaries between oscillation and dispersionless
zones.

This construction may appear to be complex and, perhaps, just as
involved as solving the original evolution equation.  If one seeks the
complete description, this is a challenging problem indeed.  However,
often one is only interested in the macroscopic physical DSW
properties encapsulated by the locations (DSW speeds) and limiting
behavior (harmonic or soliton) of the free boundaries.  For Riemann
step initial data, a DSW fitting method is available under modest
assumptions \cite{el_resolution_2005,el_dispersive_2016}.  Remarkably,
the DSW speeds and edge properties can be found by integrating two
ordinary differential equations (ODEs) that involve only the
dispersionless characteristic speeds and the dispersion relation.  The
integration of the full Whitham equations is not required.  In this
work, we utilize the DSW fitting method for oblique, spatial NLS DSWs.

We remark that this problem utilizes an unconventional application of
Whitham theory.  In typical problems, there is either a scalar or
system of two governing equations.  Here, we have three governing
equations, with one of them defining an algebraic relationship in the
dispersionless limit.

\subsection{Previous Results}
\label{sec:previous-results}

One scenario leading to dispersive hydrodynamic singularity formation
is a large disturbance, say, in the fluid density.  Through a process
of self-steepening and dispersive regularization, the disturbance
results in unsteady DSWs, which are typically realized along one
spatial dimension.  The analytical description of DSWs was pioneered
by Gurevich and Pitaevskii \cite{gurevich_nonstationary_1974} through
the use of Whitham averaging theory
\cite{whitham_non-linear_1965,whitham_linear_1974}.  (2+0)D steady
oblique DSWs can arise in the long time limit of supersonic flow past
a corner.  In the weakly nonlinear regime
\cite{gurevich_supersonic_1995,gurevich_supersonic_1996,hoefer_dark_2012},
a Korteweg-de Vries (KdV) equation describes the behavior.  In the
hypersonic \cite{el_two-dimensional_2009} regime, hypersonic
similitude was used to reduce a (2+0)D steady corner flow problem to a
(1+1)D unsteady piston problem
\cite{hoefer_piston_2008,kamchatnov_flow_2010}.  There are also (2+1)D
generalizations of (1+1)D DSWs to unsteady oblique DSWs realized by a
coordinate rotation \cite{hoefer_theory_2009,hoefer_dark_2012}.  All of
these works, except numerical simulations in
\cite{kartashov_two-dimensional_2013}, involved integrable equations
(KdV or (1+1)D NLS) enabling a detailed analytical description via the
existence of Riemann invariants for the modulation equations.  Whitham
theory is applicable to a wider class of equations and a description
of salient DSW features is possible by making a simple wave assumption
\cite{el_resolution_2005}.  We use this simple wave DSW theory to
construct large amplitude oblique DSWs and then investigate their
stability in the context of supersonic corner flow.

The layout of the manuscript is as follows.
\Cref{sec:problem-formulation} provides the formulation of the oblique
DSW problem.  We develop the necessary tools to apply Whitham theory
to the (2+0)D NLS equation in \cref{sec:prop-nls-equat}.  The
asymptotic construction of spatial oblique DSWs is completed in
\cref{sec:simple-wave-crit,sec:harmonic-edge-angle,sec:solitary-wave-edge}
and compared to known small amplitude and hypersonic regimes in
\cref{sec:small-ampl-regime,sec:hypersonic-regime}.  The properties of
spatial oblique DSWs are then studied in \cref{sec:simple-oblique-dsw}
while their admissibility and instability are studied numerically and
analytically in \cref{sec:admissibility}.  We conclude the manuscript
with some discussion in \cref{sec:disc-concl}.

\section{Problem formulation}
\label{sec:problem-formulation}

We consider the cubic defocusing or repulsive NLS equation in two
spatial dimensions
\begin{equation}
  \label{eq:1}
  i \Psi_t = - \frac{1}{2} (\Psi_{xx} + \Psi_{yy}) + | \Psi |^2 \Psi .
\end{equation}
This equation is a model equation for the order parameter of a
Bose-Einstein condensate \cite{pitaevskii_bose-einstein_2003} or the
envelope of the optical field propagating through a nonlinear medium
\cite{boyd_nonlinear_2013}.  An equivalent representation of the NLS
equation can be had through the transformation to dispersive
hydrodynamic form ($\Psi = \sqrt{\rho} e^{i \phi}$) so that equation
\cref{eq:1} becomes
\begin{subequations}
  \label{eq:58}
  \begin{align}
    \label{eq:8}
    \rho_t + (\rho \phi_x )_x + (\rho \phi_y )_y = 0  \\
    \label{eq:5}
    \phi_t + \frac{1}{2} |\nabla \phi|^2 + \rho = D[\rho], \\
    \label{eq:10}
    D[\rho] = \frac{1}{4} \left ( \frac{\rho_{xx} + \rho_{yy}}{\rho} -
      \frac{\rho_x^2 + \rho_y^2}{2 \rho^2} \right ) . 
  \end{align}
\end{subequations}
By identifying the phase gradient with a velocity field $\uv = (u,v) =
\nabla \phi$ and $\rho$ as a density, equation \cref{eq:8} expresses
the conservation of mass while equation \cref{eq:5} is an analogue of
Bernoulli's equation.  Taking the gradient of equation \cref{eq:5},
we obtain the velocity equations
\begin{subequations}
\begin{align}
  \label{eq:2}
  u_t + u u_x + v u_y + \rho_x &= D[\rho]_x \\
  \label{eq:6}
  v_t + u v_x + v v_y + \rho_y &= D[\rho]_y .
\end{align}
\end{subequations}
By virtue of potential flow, we also have the irrotationality
constraint
\begin{equation}
  \label{eq:17}
  v_x - u_y = 0 .
\end{equation}
Based on this hydrodynamic representation, it is natural to define the
\textit{dynamic} sound speed $c$ as the speed at which long
wavelength, small amplitude density perturbations propagate $c(\rho) =
\sqrt{\rho}$.  Similarly, we define the Mach number of the flow as
\begin{equation}
  \label{eq:63}
  M = \frac{|\uv|}{\sqrt{\rho}} ,
\end{equation}
and the associated Mach angle
\begin{equation}
  \label{eq:9}
  \mu = \textrm{arcsin}(M^{-1}) .
\end{equation}
We will move freely between the Cartesian velocity components $u$, $v$
and the Mach number $M$ or angle $\mu$ and flow direction
\begin{equation}
  \label{eq:12}
  \theta = \tan^{-1} (v/u),
\end{equation}
coordinates.

We are interested in stationary solutions of equation \cref{eq:58} in
the form
\begin{equation}
  \label{eq:3}
  \rho(x,y,t) \to \rho(x,y), \quad \phi(x,y,t) \to -\epsilon t + \phi(x,y),
  \quad \uv(x,y,t) \to \uv(x,y),
\end{equation}
where the constant $\epsilon$ can be interpreted as the superfluid chemical
potential \cite{pitaevskii_bose-einstein_2003} or the optical propagation
constant \cite{boyd_nonlinear_2013}.  Then equation \cref{eq:5}
becomes
\begin{align}
  \label{eq:4}
  \frac{1}{2}(u^2 + v^2) + \rho - D[\rho] &= \epsilon ,
\end{align}
an analog of Bernoulli's equation, which implicitly determines $\rho$
in terms of the flow speed $|\mathbf{u}|^2$.  Equation \cref{eq:4},
the steady continuity equation \cref{eq:8}
\begin{equation}
  \label{eq:11}
  (\rho u)_x + (\rho v)_y = 0,
\end{equation}
and the irrotationality constraint \cref{eq:17} constitute a closed
system for $\rho$, $u$, and $v$.  With some algebraic manipulation,
this system can be written in the form
\begin{subequations}
  \label{eq:62}
  \begin{align}
    \label{eq:59}
    \begin{bmatrix}
      u^2 - \rho & uv \\
      0 & 1
    \end{bmatrix}
    \begin{bmatrix}
      u \\
      v
    \end{bmatrix}_x + 
    \begin{bmatrix}
      uv & v^2 - \rho \\
      -1 & 0
    \end{bmatrix}
    \begin{bmatrix}
      u \\
      v
    \end{bmatrix}_y &= 
    \begin{bmatrix}
      \uv \cdot \nabla D[\rho] \\
      0
    \end{bmatrix} , \\
    \label{eq:61}
    \rho &= \epsilon - \frac{1}{2} | \uv |^2  + D[\rho] .
  \end{align}
\end{subequations}

\subsection{Corner Boundary Value Problem}
\label{sec:corn-bound-value}

\begin{figure}
  \centering
  \includegraphics{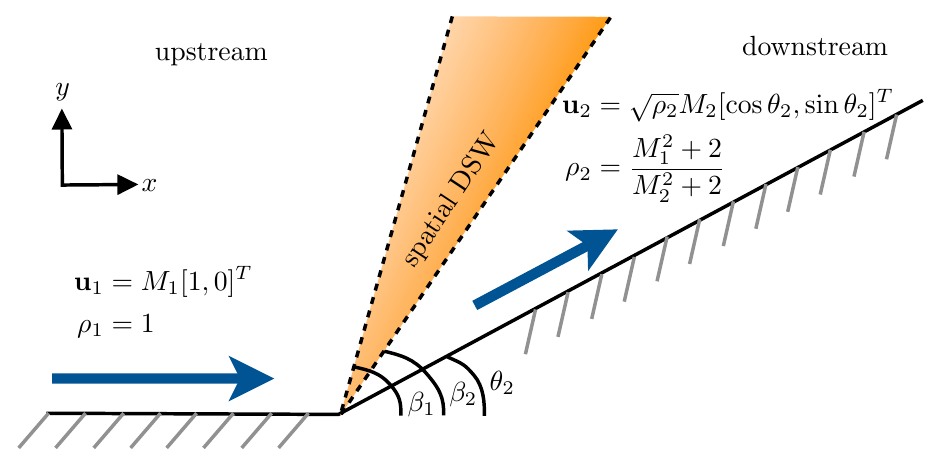}
  \caption{Corner boundary value problem for steady, supersonic NLS flow.}
  \label{fig:bvp}
\end{figure}

The steady equations \cref{eq:62} are supplemented with boundary
conditions appropriate for supersonic flow past a corner.  
\Cref{fig:bvp} provides a schematic for the flow of interest.  Without
loss of generality, we scale the upstream density to unity and
consider an incoming, uniform flow parallel to the $x$ axis with Mach
angle $\mu_1$.  Evaluating equation \cref{eq:61} with these far field
conditions as $x \to -\infty$ gives
\begin{equation}
  \label{eq:64}
  \epsilon = 1 + \csc^2 \mu_1/2.
\end{equation}
At all physical boundaries, we apply the condition
\begin{equation}
  \label{eq:65}
  \uv \cdot \mathbf{n} = 0,
\end{equation}
where $\mathbf{n}$ is normal to the boundary.  Subsequently, the
downstream flow points in the direction of the corner angle $\theta_2$
with Mach angle $\mu_2$ and density $\rho_2$.  In the far field
downstream flow, we evaluate equation \cref{eq:61} again with
\cref{eq:64} to find
\begin{equation}
  \label{eq:66}
  \rho_2 = \frac{M_1^2 + 2}{M_2^2 + 2} = \frac{\sin^2\mu_2(1 +
    2\sin^2\mu_1)}{\sin^2 \mu_1(1 + 2\sin^2\mu_2)} . 
\end{equation}
We see that knowledge of the fluid velocity in a steady configuration
determines the fluid density via the generalized Bernoulli's equation
\cref{eq:61} as in classical gas dynamics.

Because the upstream and downstream flows exhibit a jump at $x = 0$ in
the density and velocity field, we can view this steady configuration
as a dispersive Riemann problem with ``initial'' conditions
\begin{equation}
  \label{eq:67}
  \mu(x,0) = 
    \begin{cases}
      \mu_1 & x < 0 \\
      \mu_2 & x > 0
    \end{cases}, \quad 
    \theta(x,0) = 
    \begin{cases}
      0 & x < 0 \\
      \theta_2 & x > 0
    \end{cases}, \quad \rho(x,0) = \begin{cases}
      1 & x < 0 \\
      \rho_2 & x > 0
    \end{cases} ,
\end{equation}
where $\rho_2$ is given by \cref{eq:66}.  We can treat the spatial
coordinate $y$ as a time-like variable.  As such, we will often use
common terminology related to dynamical behavior but the reader is
advised to keep in mind that we are considering steady, i.e.,
time-independent flow configurations.  The Riemann problem
\cref{eq:67} for equations \cref{eq:62} differs from the classical
Riemann problem in that a dispersive regularization is
required---i.e., we need to incorporate the dispersive effects
encapsulated in the term $D[\rho]$, equation \cref{eq:10}---in
contrast to a dissipative regularization.  Generically, the Riemann
problem for systems is resolved into constant states connected by
multiple waves, each of which can be a rarefaction or shock wave
\cite{lax_hyperbolic_1973}.  Motivated by the geometry of the problem,
it is natural to consider the case where the upstream Mach angle
$\mu_1$ and deflection angle $\theta_2$ are given so that $\mu_2$ is
to be determined so that the Riemann problem results in a single wave,
a simple DSW.  This is demonstrated in \cref{sec:oblique-dsws}.
Furthermore, we will show in \cref{sec:disp-regime} that a necessary
condition for the generation of a DSW requires supersonic upstream
flow $M_1 > 1$ ($0 < \mu_1 < \pi/2$) and a compressive turn $0 <
\theta_2 < \pi/2$.  The spatial oblique DSW will be shown to exhibit
modulated, periodic waves inside a wedge with bounding angles
$\beta_1$ and $\beta_2$ emanating from the corner (see schematic in
\cref{fig:bvp}).

Stated concisely, we seek a large $y$ description of the dispersive
Riemann problem \cref{eq:67} for equations \cref{eq:62} that results
in a single spatial oblique DSW.

\section{Properties of the stationary NLS equation}
\label{sec:prop-nls-equat}

The application of DSW fitting theory requires knowledge of certain
properties of the governing equation \cite{el_dispersive_2016}.  In
this section, we present the needed properties.

\subsection{Spatially Periodic Solution}
\label{sec:peri-trav-wave}

Equations \cref{eq:62} admit the periodic wave solution
\cite{el_two-dimensional_2007} 
\begin{subequations}
  \label{eq:49}
  \begin{align}
    \label{eq:29}
    \rho(x,y) &= \rho(\xi ) = p_1 + (p_2 - p_1) \,
    \mathrm{sn}^2\left (\sin \varphi \sqrt{p_3 - p_1}\xi;m \right), \\
    \label{eq:50}
    u(x,y) &= d \cos \varphi + \sigma \frac{\sin \varphi \sqrt{p_1 p_2
      p_3}}{\rho(\xi)}, \\
    \label{eq:51}
     v(x,y) &= d \sin \varphi - \sigma \frac{\cos \varphi \sqrt{p_1 p_2
      p_3}}{\rho(\xi)}, \quad \xi = x -
    \cot \varphi \, y , \quad m = \frac{p_2 - p_1}{p_3 - p_1},
  \end{align}
\end{subequations}
where $\mathrm{sn}$ is the Jacobi elliptic function.  This periodic
solution has five parameters $0 < p_1 < p_2 < p_3$, $d \in \R$, and
$\varphi \in (0,\pi/2)$.  Both choices of the sign $\sigma = \pm 1$ give
valid solutions corresponding to waves propagating in opposite
directions.  The angle of constant phase $\xi = \mathrm{const}$ is
$\varphi$, measured from the $x$ axis.  Recalling that $y$ can be viewed
as a time-like variable, $\cot \varphi$ is the ``speed'' of propagation
whereas $\tan \varphi$ is the slope of the wavefronts (lines of constant
phase) in the $x$-$y$ plane.  Here, $d$ is a free parameter whose
physical meaning is the uniform flow velocity along the phase
wavefronts; this flow is normal to the direction of wave
``propagation'' and therefore it does not affect the nonlinear wave
profile $\rho(x,y)$.  The period of the wave in the $x$ direction is
\begin{equation}
  \label{eq:25}
  L = \frac{2 K(m)}{\sin \varphi \sqrt{p_3 - p_1}},
\end{equation}
where $K(m)$ is the complete elliptic integral of the first kind.  We
remark that the oblique periodic wave \cref{eq:49} was generated by
simply rotating the zero phase speed, spatially periodic solution of
the (1+1)D NLS equation with the addition of an arbitrary flow along
the phase wavefronts.

It is convenient to consider a more physically inspired set of
parameters that uniquely determine the periodic wave.  They are the
wavenumber in the $x$ direction $k = 2\pi /L$, the density oscillation
amplitude $a = p_2 - p_1$, and the mean values $\orho$,
$\ou$, $\ov$ computed as
\begin{equation}
  \label{eq:7}
  \overline{f} \equiv \frac{1}{L} \int_0^L f(\xi) \ud \xi ,
\end{equation}
leading to
\begin{align}
  \label{eq:13}
  \orho &= p_3 - (p_3 - p_1) \frac{E(m)}{K(m)}, \\
  \ou &= d \cos \varphi + \sigma \frac{\sin \varphi \, p_2 p_3}{p_1 K(m)}
  \Pi(1-p_2/p_1,m), \\
  \ov &= d \sin \varphi - \sigma \frac{\cos \varphi\, p_2 p_3}{p_1 K(m)}
  \Pi(1-p_2/p_1,m) ,
\end{align}
where $E(m)$ and $\Pi(1-p_2/p_1,m)$ are the complete elliptic
integrals of the second and third kinds, respectively.  An auxiliary
variable, the oscillation ``frequency'', can be defined as $\omega = k
\cot \varphi$.

Dispersive shock waves exhibit two distinct edges: a harmonic wave
edge exhibiting small amplitude oscillations and a solitary wave edge.
These two features are captured by the solution \cref{eq:49} for
appropriate limiting cases \cite{el_two-dimensional_2007}.  The $k \to
0$ limit yields the stationary oblique soliton solution
\cite{el_oblique_2006}
\begin{subequations}
  \label{eq:52}
  \begin{align}
    \label{eq:30}
    \rho(x,y) &= \rho(\zeta) = \orho - a \,
      \mathrm{sech}^2(\zeta), \\
    \label{eq:54}
    u(x,y) &= \orho^{1/2} \overline{M}\left (
      \cos (\varphi-\oth) \cos
      \varphi + \sin (\varphi-\oth) \sin \varphi\,
      \frac{\orho}{\rho(\zeta)} \right ), \\
    \label{eq:55}
    v(x,y) &= \orho^{1/2} \overline{M}\left ( \cos
      (\varphi-\oth) \sin \varphi -\sin
      (\varphi-\oth) \cos \varphi
      \frac{\orho}{\rho(\zeta)}
    \right ), \\
    \label{eq:16}
    \zeta &= a^{1/2} \sin \varphi (x - \cot \varphi \,y),
  \end{align}
  where $\overline{\uv} = \orho^{1/2} \overline{M} [\cos \oth,\sin
  \oth]^T = [\ou,\ov]^T$ is the far field flow with $\overline{M}$,
  $\oth$ the averaged Mach number \cref{eq:63} and averaged flow
  angle \cref{eq:12}.  The amplitude-slope relation
  \begin{equation}
    \label{eq:15}
    a = \orho\left ( 1 - \overline{M}^2 \sin^2(\varphi-\oth) \right ), 
  \end{equation}
\end{subequations}
and the  positive density restriction $0 < a$ imply
\begin{equation}
  \label{eq:31}
  |\sin (\varphi - \oth) | < \sin \omu,
\end{equation}
so that the oblique solitary wave is oriented inside the Mach cone
defined by the Mach angle $\omu$.  In contrast, small amplitude
dispersive waves exhibit lines of constant phase oriented outside the
Mach cone (shown below).  The Mach cone is well-known from gas
dynamics and represents the spatial region where small amplitude
disturbances are confined to propagate in dispersionless, supersonic
flow (cf.~\cite{courant_supersonic_1948}).  This does not contradict
the orientation of small amplitude dispersive waves because their
propagation direction is orthogonal to the direction of constant
phase, hence is outside the Mach cone.  These restrictions on wave
orientation have been previously described in the context of
supersonic NLS flow past a small impurity where oblique solitary waves
and dispersive waves are generated \cite{el_two-dimensional_2007}.

A calculation shows that the $a \to 0$ limit in \cref{eq:49}
corresponds to harmonic waves satisfying
\begin{align}
  \label{eq:24}
  \rho(x,y) &= \orho - \frac{a}{2} \cos (kx - \omega y) +
  \mathcal{O}(a^2), \\
  u(x,y) &= \ou + a \frac{\sigma k}{4 \orho} \left
    ( 1 + \frac{4 \orho}{k^2 + \omega^2} \right )^{1/2} \cos(kx - \omega y) +
  \mathcal{O}(a^2), \\
  v(x,y) &= \ov - a \frac{\sigma \omega}{4 \orho} \left
    ( 1 + \frac{4 \orho}{k^2 + \omega^2} \right )^{1/2} \cos(kx - \omega y) +
  \mathcal{O}(a^2) ,
\end{align}
with the dispersion relation

\begin{equation}
  \label{eq:69}
  \orho(k^2 + \omega^2) \left ( 1 + \frac{k^2 + \omega^2}{4
      \orho} \right ) - (k\ou - \omega \ov)^2 = 0 .
\end{equation}
The long wave, \textit{stationary} sound speed or inverse slope
$\lambda = \lim_{k \to 0} \omega/k$ can be found by dividing equation
\cref{eq:69} by $k^2$ and taking the limit $k \to 0$, leading to the
relation
\begin{equation}
  \label{eq:19}
  \orho(1 + \lambda^2) - (\ou - \lambda \ov)^2 = 0,
\end{equation}
whose roots are
\begin{equation}
  \label{eq:85}
  \begin{split}
    \lambda_{\pm} &= \frac{\ou\, \ov \mp \sqrt{(\ou^2 + \ov^2 -
        \orho)\orho }}{\ov^2 -
      \orho}  = \cot \left ( \oth \pm \omu \right ) .
  \end{split}
\end{equation}
Thus, as $y$ increases, long stationary waves on the flow
$(\orho,\overline{\uv})$ exhibit the ``slopes of sound''
$1/\lambda_{\pm}$.  Real sound slopes coincide with supersonic flow
($\overline{M} > 1$) and modulationally stable waves, consistent with
hyperbolicity of the dispersionless equations (see
\cref{sec:disp-regime}).  Complex sound slopes correspond to
subsonic flow and modulational instability, the counterpart of
ellipticity for the dispersionless equations.

\begin{figure}
  \centering
  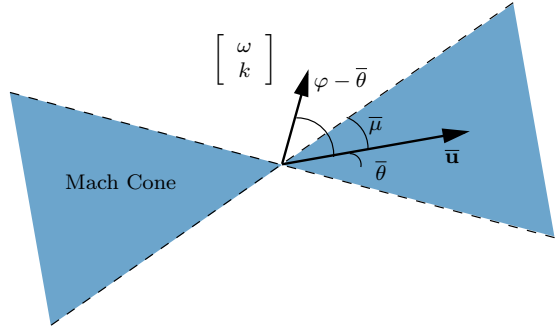
  \caption{Admissible vectors $[\omega,k]^T = q[\cos \varphi, \sin
    \varphi]^T$ for small amplitude dispersive waves relative to the
    background flow direction $\overline{\mathbf{u}}$.  The shaded
    region depicts the Mach cone of inadmissible directions defined by
    the Mach angle $\overline{\mu} = \sin^{-1} (\overline{M}^{-1})$.}
  \label{fig:disp_relation_schematic}
\end{figure}
It is convenient to consider the dispersion relation \cref{eq:69} in
polar coordinates.  For this we follow \cite{gladush_radiation_2007}
and take
\begin{equation}
  \label{eq:70}
  k = q \sin \varphi, \quad \omega = q \cos \varphi, \quad \overline{\uv} =
  \orho^{1/2} \overline{M}[\cos \oth, \sin \oth] .
\end{equation}
Then equation \cref{eq:69} becomes
\begin{equation}
  \label{eq:71}
  q = 2 \orho^{1/2} \left ( \frac{\sin^2 (\varphi - \oth)}{\sin^2 \omu}
    - 1 \right )^{1/2} .
\end{equation}
The dispersion relation \cref{eq:71} is real-valued so long as
\begin{equation}
  \label{eq:72}
  |\sin(\varphi - \oth)| \ge \sin \omu .
\end{equation}
\begin{figure}
  \centering
  \includegraphics[width=0.62\columnwidth,clip=true,trim=0 185 0 185]{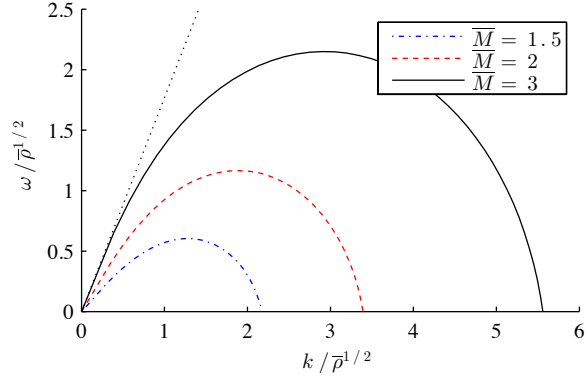}
  \caption{The dispersion relation for several Mach numbers
    $\overline{M}$ and flow angle $\oth = \pi/18$.  The dotted line
    corresponds to the boundary of the Mach cone for $\overline{M} =
    3$.}
  \label{fig:disp_relation}
\end{figure}
This inequality can be interpreted in the following way.  The
direction of constant phase $[\omega,k]^{T}$, parallel to the
dispersive wave troughs and crests, has angle $\varphi$.  Inequality
\cref{eq:72} is satisfied when $\varphi$ is \emph{outside} the Mach cone
defined by the static sound slope angles $\oth \pm \omu$ and
$\oth \pm \omu + \pi$ shown in 
\cref{fig:disp_relation_schematic}.  Hence, inequality \cref{eq:72}
leads to the admissible angles satisfying
\begin{equation}
  \label{eq:73}
  \omu  \le  \left | \varphi - \oth \right | \le \pi -
  \omu ,
\end{equation}
in which small amplitude waves satisfying the dispersion relation
\cref{eq:69} can propagate.  This admissible region is depicted in
\cref{fig:disp_relation_schematic}.  A branch of the dispersion
relation \cref{eq:69} in the $\omega$-$k$ plane for specific choices
of $\overline{M}$ is plotted in \cref{fig:disp_relation}.  This
primary branch corresponds to $\varphi \in [\oth + \omu,\oth + \pi/2]$.
The dispersion relation is concave (negative dispersion) across the
band $[0,k_{\rm max}]$ of admissible wavenumbers where $k_{\rm max}$
is reached when $\varphi = \pi/2$.  The group velocity is
\begin{equation}
  \label{eq:47}
  \omega_k = \frac{\omega_\varphi}{k_\varphi} = \frac{\sin(2 \oth - 3 \varphi)
    + \cos 2\omu \sin \varphi}{\cos (2 \oth - 3 \varphi) - \cos 2 \mu \cos
    \varphi} ,
\end{equation}
which reaches a zero point for $k_0 \in (0,k_{\rm max})$.  

\subsection{Whitham Modulation Equations}
\label{sec:conservation-laws}

Whitham averaging theory describes oscillatory regions by allowing the
nonlinear periodic wave solution's parameters to vary slowly relative
to the period of oscillation.  The modulation equations can be
determined in several, equivalent ways
\cite{whitham_non-linear_1965,whitham_linear_1974}.  In DSW theory, it
is common to obtain the modulation equations by averaging conservation
laws associated to the evolution equation with the periodic wave
\cite{el_dispersive_2016}.  For this, we allow the periodic solution's
\cref{eq:49} five parameters to vary according to $(\ou,\ov,\orho,a,k)
= (\ou,\ov,\orho,a,k)(x,y)$.  Due to slow variation, any derivatives
of the parameters with respect to $x$ or $y$ are considered
asymptotically smaller than differentiation of the periodic wave with
respect to its phase $\xi$.  We begin by averaging equation
\cref{eq:4} subject to the periodic wave solution \cref{eq:49}
\begin{equation}
  \label{eq:18}
  \frac{1}{2}(\overline{u^2} + \overline{v^2}) + \orho -
  \overline{D[\rho]} = \epsilon ,
\end{equation}
recalling the definition of nonlinear wave averaging \cref{eq:7}.
This provides a global relation for the modulation parameters, thus
reducing the number of required conservation laws by one.  Because of
this, we can view the mean density $\orho$ as an auxiliary
variable, determined from \cref{eq:18} by the variations of the other
parameters.  Three more equations are found by averaging the
continuity equation \cref{eq:11}, the irrotationality constraint
\cref{eq:17}, and the stationary energy equation
\cite{jin_semiclassical_1999}
\begin{subequations}
  \label{eq:23}
  \begin{align}
    \label{eq:20}
    \nabla \cdot (\overline{\rho \uv}) &= 0, \\
    \label{eq:21}
    (\ov)_x + (\ou)_y &= 0, \\
    \label{eq:56}
    \nabla \cdot \left ( \overline{\uv \left ( \mathcal{E} + \rho^2/2 \right )} 
      - \overline{\uv \Delta \rho/4} + \overline{(\nabla \rho) (\nabla \cdot (\rho
        \uv))/(4\rho)} \right ) &= 0 .
  \end{align}
\end{subequations}
where the energy $\mathcal{E}$ is
\begin{equation}
  \label{eq:57}
  \mathcal{E} = \frac{1}{2} \rho (u^2 + v^2) + \frac{| \nabla \rho
    |^2}{8 \rho} + \frac{1}{2} \rho^2 .
\end{equation}
The final modulation equation is the conservation of waves,
\begin{equation}
  \label{eq:22}
  k_y + \omega_x = 0,
\end{equation}
a consistency condition for the assumption of slow modulations
\cite{whitham_linear_1974}.

The modulation equations \cref{eq:18}, \cref{eq:23}, and
\cref{eq:22} provide a closed system for the spatial evolution of the
periodic wave's five parameters.  In order to construct the DSW, we
need to assign appropriate initial/boundary data to the modulation
equations compatible with the Riemann data \cref{eq:67}.  This will
be achieved in \cref{sec:oblique-dsws}.

\subsection{Dispersionless Regime}
\label{sec:disp-regime}

The application of DSW theory requires knowledge of the dispersionless
limiting equations ($D \to 0$ in equations \cref{eq:62}), which we
can write as
\begin{subequations}
  \label{eq:14}
  \begin{align}
  \label{eq:26}
    \begin{bmatrix}
      u \\[3mm]
      v
    \end{bmatrix}_y +
    \begin{bmatrix}
      0 & -1 \\
      \frac{u^2 - \rho}{v^2 - \rho} & \frac{2 uv}{v^2 - \rho}
    \end{bmatrix}
    \begin{bmatrix}
      u \\[3mm]
      v
    \end{bmatrix}_x &= 0 , \\
    \label{eq:27}
    \rho + \frac{1}{2} |\uv|^2 &= \epsilon .
  \end{align}
\end{subequations}
This quasi-linear system has been well-studied in the context of gas
dynamics (cf.~\cite{courant_supersonic_1948}).  It exhibits the same
characteristic speeds $\lambda_\pm = \cot (\theta \pm \mu)$ as
resulted from the long wavelength limit \cref{eq:85}.  The system
\cref{eq:26} is strictly hyperbolic and genuinely nonlinear when the
characteristic speeds are real, i.e., when the flow is supersonic.
Equations \cref{eq:26} are diagonalized by the Riemann invariants
\begin{equation}
  \label{eq:28}
  s_\pm(\theta,\mu) = \theta \pm \left ( - \mu + \sqrt{3} \tan^{-1} ( \sqrt{3}
    \tan \mu ) \right ) .
\end{equation}

Simple wave solutions correspond to changes in only one Riemann
invariant
\begin{equation}
  \label{eq:53}
  s_\pm = \mathrm{const} \quad \mathrm{and} \quad \lambda_{\mp} =
  x/y . 
\end{equation}
These solutions exhibit flows with an expansion turn forming a
rarefaction wave.  The Riemann initial data \cref{eq:67} for the
dispersionless equations \cref{eq:26} exhibits a single rarefaction
wave when
\begin{equation}
  \label{eq:60}
  s_-(0,\mu_1) < \theta_2 < 0.
\end{equation}
This is the Prandtl-Meyer expansion fan
\cite{courant_supersonic_1948}.  When the flow turns too much,
$\theta_2 < s_-(0,\mu_1)$, it exhibits cavitation ($\rho \to 0$).

The dispersionless equations \cref{eq:14} exhibit gradient
catastrophe when the flow experiences a compressive turn.  For the
Riemann data \cref{eq:67}, this corresponds to $\theta_2 > 0$.  In
gas dynamics, the singularity is resolved by appealing to small scale
physical processes coinciding with dissipation.  In the next section,
we resolve singularity formation utilizing a dispersive regularization
due to nonzero $D[\rho]$ in equations \cref{eq:62}.

\section{Oblique DSWs}
\label{sec:oblique-dsws}

As described in \cref{sec:simple-dsws}, a complete description of
spatial DSWs requires integration of the Whitham equations
\cref{eq:18}, \cref{eq:23}, and \cref{eq:22} matched to the
dispersionless flow satisfying \cref{eq:14}.  As originally
formulated by Gurevich and Pitaevskii in the KdV DSW problem
\cite{gurevich_nonstationary_1974}, matching corresponds to equality
of the dispersionless $(u,v,\rho)$ and averaged $(\ou,\ov,\orho)$ flow
variables at the edges of the DSW.  Embedded within this matching
procedure is the requirement that two characteristics of the Whitham
equations coalesce, which occurs when either $a \to 0$ (harmonic wave
edge) or $k \to 0$ (solitary wave edge).  The remaining modulation
parameter determines the location of the interface.  At the harmonic
wave edge, the nonzero wavenumber corresponds to a wave packet moving
with the group velocity.  The solitary wave edge moves according to
the phase speed of a solitary wave with amplitude $a$.  The specific
choice of Riemann data \cref{eq:67} implies self-similar behavior for
the modulations so it is natural to seek a simple wave solution of the
modulation equations.  The innovation in \cite{el_resolution_2005} was
the determination of the DSW edge locations while bypassing full
integration of the modulation equations.  This was achieved by a
priori assuming the existence of a simple wave and analyzing the
modulation equations in the $a \to 0$ and $k \to 0$ limits.  This
analysis results in a universal description of simple DSW edge speeds.

In \cref{sec:prop-nls-equat} we provided all the necessary pieces in
order to implement Whitham averaging.  However, it is important to
note that the simple DSW construction presented in
\cite{el_resolution_2005} can be implemented with knowledge of only
the linear dispersion relation \cref{eq:69}, the dispersionless
characteristic speeds \cref{eq:85}, and the solitary wave
amplitude/speed relation \cref{eq:15}.  This reflects the fact that a
DSW incorporates a range of nonlinear wave phenomena, from large
amplitude solitary waves to vanishingly small amplitude waves, into a
single coherent structure.  The integration of the universal, simple
DSW ODEs show how the harmonic and solitary wave edges are nonlocally
connected.
 
\subsection{DSW Locus}
\label{sec:simple-wave-crit}

The assumption of a simple wave solution to the full modulation system
yields several important results.  Utilizing a backward characteristic
argument, it was shown that the Riemann data \cref{eq:67} coincides
with a single DSW opening to the right when \cite{el_resolution_2005}
\begin{equation}
  \label{eq:77}
  s_-(0,\mu_1) = s_-(\theta_2,\mu_2) .
\end{equation}
This is a 2-DSW, named thus because it degenerates to the faster
characteristic speed $\lambda_+$ in the small amplitude regime
\cite{el_dispersive_2016}.  This provides an explicit relationship
between the upstream and downstream flow variables that must hold for
a 2-DSW.  In combination with two expansion fan solutions and 1-DSWs,
these four wave types form the generic building blocks for the general
solution to the Riemann problem.  See \cite{el_dispersive_2016} for
additional details.  The locus \cref{eq:77} enables the determination
of the \textit{sonic curve}, i.e., the relationship between $\mu_1$
and $\theta_2$ such that $M_2 = 1$ or, equivalently, $\mu_2 = \pi/2$.
We define the sonic angle $\theta_{\rm s}$ according to $s_-(0,\mu_1)
= \lim_{\mu_2 \to \pi/2^{-}} s_-(\theta_{\rm s},\mu_2)$, obtaining
\begin{equation}
  \label{eq:97}
  \theta_{\rm s}(\mu_1) = \left ( \sqrt{3}-1 \right )\frac{\pi}{2} +
  \mu_1 - \sqrt{3} \tan^{-1} \left ( \sqrt{3} \tan \mu_1 \right ) .
\end{equation}
For an oblique DSW with $0 < \mu_1 < \pi/2$, the dispersionless
quasi-linear system \cref{eq:26} is hyperbolic if and only if $0 \le
\theta_2 < \theta_{\rm s}$.

In addition to the DSW locus \cref{eq:77}, a local, simple wave
relation holds when either $a = 0$ or $k = 0$
\cite{el_resolution_2005,el_dispersive_2016}
\begin{equation}
  \label{eq:75}
  s_-(0,\mu_1) = s_-(\oth,\omu) , \quad 0 <
  \oth(\omu) < \theta_2 ,
\end{equation}
leading to
\begin{equation}
  \label{eq:84}
  \oth = \oth(\omu) = \mu_1 - \omu - \sqrt{3} \left [
    \tan^{-1}(\sqrt{3} \tan \mu_1 ) - \tan^{-1}(\sqrt{3}\tan \omu)
  \right ] .
\end{equation}
Similarly, when $a = 0$ or $k = 0$
equation \cref{eq:18} implies the local relation
\begin{equation}
  \label{eq:80}
  \orho = \orho(\omu) = \frac{\sin^2 \omu\, (1 + 2 \sin^2 \mu_1)}{\sin^2
    \mu_1 \, (1 + 2 \sin^2 \omu)} .
\end{equation}
We can evaluate the density between the oblique DSW and the wedge as
\begin{equation}
  \label{eq:86}
  \rho_2 = \orho(\mu_2) = \frac{\sin^2 \mu_2(1 + 2 \sin^2
    \mu_1)}{\sin^2 \mu_1(1 + 2 \sin^2 \mu_2)} .
\end{equation}

\subsection{Harmonic Wave Edge Angle}
\label{sec:harmonic-edge-angle}

The angle $\beta_1$ (recall \cref{fig:bvp}) of the harmonic wave
edge ($a = 0$) is determined by integrating the simple wave ordinary
differential equation \cite{el_resolution_2005}
\begin{equation}
  \label{eq:74}
  \frac{\mathrm{d} k}{\mathrm{d} \omu} = \frac{\omega_{\omu}
    (k,\omu)}{\lambda_+(\omu) - \omega_k(k,\omu)}, \quad k(\mu_2) = 0
  , 
\end{equation}
to $k_1 = k(\mu_1)$ and then evaluating the group velocity
\begin{equation}
  \label{eq:34}
  \beta_1 = \cot^{-1} \omega_k(k_1,\mu_1).
\end{equation}
This represents an integration from the solitary wave edge
$(\omu,\oth,\orho,k,a) =$ $(\mu_2,\theta_2,\rho_2,0,a_2)$ to the
harmonic wave edge $(\omu,\oth,\orho,k,a) = (\mu_1,0,1,k_1,0)$.  The
integration is accomplished while remaining in the $a = 0$ plane.  Let
us see why this is so.  The assumed existence of a simple wave
necessitates a functional relationship between the modulation
variables $F(\omu,\oth,\orho,k,a) = C$ where $C \in \R$ is a constant.
Then, $C$ is implicitly determined by $F(\mu_2,\theta_2,\rho_2,0,a_2)
= C$ from the soliton edge initial condition $k(\mu_2) = 0$ when $\omu
= \mu_2$, $\oth = \theta_2$, and $\orho = \rho_2$.  But the initial
condition is independent of $a$ so the functional relationship
$F(\mu_2,\theta_2,\rho_2,0,0) = C$ must also hold and determines $C$.
Then $k_1$ is found by evaluating $F(\mu_1,0,1,k_1,0) = C$
via integration of \cref{eq:74} in the $a = 0$ plane, thus bypassing
the integration of the full Whitham system.

Due to the implicit nature of the dispersion relation \cref{eq:69}, we
undertake the solution of \cref{eq:74} utilizing the polar form
\cref{eq:70}, \cref{eq:71}.  The transformation of \cref{eq:74} into
polar form is somewhat involved.  We relegate the details to
\cref{sec:harmonic-edge}.  The harmonic edge initial value problem
then becomes
\begin{equation}
  \label{eq:91}
  \frac{\mathrm{d} \varphi}{\mathrm{d} \omu} = \frac{2 \cos \omu \left [
        2 \sin \omu - \sin (2\oth+\omu-2\varphi)  
      \right ]}{(2-\cos 2\omu) \left [ \sin (2\omu) - 2 \sin(2 \oth-2\varphi)
      \right ]} , \quad \varphi(\mu_2) = \oth(\mu_2) + \mu_2 ,
\end{equation}
where $\oth(\omu)$ satisfies the local relation \cref{eq:75}.
Integrating this to $\omu = \mu_1$ and evaluating \cref{eq:34}
provides the harmonic edge angle.

\subsection{Solitary Wave Edge Angle}
\label{sec:solitary-wave-edge}

The angle $\beta_2$ (recall \cref{fig:bvp}) of the solitary wave
edge ($k=0$) can be formulated in an analogous way to the harmonic
edge by the introduction of new modulation variables $(k,\omega) \to
(\kt,\wt)$ via 
\begin{equation}
  \label{eq:33}
  \wt(\kt,\omu) = - i \omega(i \kt,\omu) .
\end{equation}
The conjugate wavenumber $\kt$ plays the role of an amplitude and is
a convenient parameterization of the periodic traveling wave, which leads
to the simple wave initial value problem \cite{el_resolution_2005}
\begin{equation}
  \label{eq:76}
  \frac{\mathrm{d} \kt}{\mathrm{d} \omu} =
  \frac{\wt_{\omu}(\kt,\omu)}{\lambda_+(\omu) - 
    \wt_{\kt}(\kt,\omu)}, \quad \kt(\mu_1) = 0 .
\end{equation}
Note the symmetry of equations \cref{eq:74} and \cref{eq:76}.  Upon
integration of \cref{eq:76}
to $\kt_2 = \kt(\mu_2)$, the solitary wave edge angle is determined
from the solitary wave phase speed $\wt/\kt$ according to
\begin{equation}
  \label{eq:32}
  \beta_2 = \cot^{-1} \frac{\wt(\kt_2,\mu_2)}{\kt_2} .
\end{equation}
The solution of equation \cref{eq:76} represents the integration from
the harmonic wave edge to the solitary wave edge in the $k = 0$
plane in an analogous manner to the harmonic edge case.  As shown in
\cref{sec:soliton-edge}, this problem can be cast in the form 
\begin{align}
  \label{eq:94}
  \frac{\mathrm{d} \phit}{\mathrm{d} \omu}
  &= \frac{2 \cos \omu \left [ 2 \sin \omu - \sin (2\oth+\omu-2\phit) 
    \right ]}{(2-\cos 2\omu) \left [ \sin (2\omu) - 2 \sin(2 \oth-2\phit)
    \right ]} , \quad
  \phit(\mu_1) = \oth(\mu_1) + \mu_1 = \mu_1 ,
\end{align}
where $\oth(\omu)$ satisfies \cref{eq:75}.  Integrating to $\omu =
\mu_2$ and evaluating \cref{eq:32} provides the soliton edge angle.

\subsection{Small Amplitude Regime}
\label{sec:small-ampl-regime}

The small amplitude regime was studied in \cite{hoefer_dark_2012} by
an asymptotic reduction of the (2+0)D NLS equation \cref{eq:62} to the
KdV equation.  This regime corresponds to small deflection angles $0 <
\theta_2 \ll 1$ with $M_1 = \mathcal{O}(1)$, to distinguish it from
the hypersonic regime studied in the next section.  Here, we perform
an asymptotic analysis of the Whitham modulation equations for the
full (2+0)D equations to verify the validity of our results and to
provide an alternative derivation.  Evaluating the DSW locus
\cref{eq:75} in this regime implies
\begin{equation}
  \label{eq:81}
  \oth(\omu) = \frac{2 (\omu - \mu_1)}{1 + 3 \tan^2 \mu_1} +
  \mathcal{O}((\omu - \mu_1)^2) ,
\end{equation}
which gives $\theta_2 = \oth(\mu_2)$.  This relation can be inverted
to give
\begin{equation}
  \label{eq:88}
  \omu(\oth) = \mu_1 + \frac{1}{2}(1 + 3 \tan^2 \mu_1) \oth +
  \mathcal{O}(\oth^2) .
\end{equation}
We also recover the asymptotic density variation
\begin{equation}
  \label{eq:83}
  \orho(\oth) = 1 + 2 \csc(2\mu_1) \oth + \mathcal{O}(\oth^2) .
\end{equation}
Both \cref{eq:88} and \cref{eq:83} agree with the results in
\cite{hoefer_dark_2012} when $\oth = \theta_2$.

Since $0 < \oth \ll 1$, the restriction \cref{eq:73} implies $\varphi -
\mu_1 = \mathcal{O}(\oth)$ in the parametric representation of the
dispersion relation.  It is convenient to utilize the independent
variable $\oth$ in the simple wave initial value problem \cref{eq:91}
rather than $\omu$, achieved through the relation \cref{eq:81}.
Then, for $0 < \oth \ll 1$, \cref{eq:91} becomes
\begin{equation}
  \label{eq:82}
  \frac{\mathrm{d}\varphi}{\mathrm{d} \oth} = \frac{1}{2} \sec^2 \mu_1,
  \quad \varphi(\theta_2) = \theta_2 + \mu_2 .
\end{equation}
The solution evaluated at the harmonic edge, $\oth = 0$, yields
$\varphi_1 \equiv \varphi(0) = \mu_2 + (1-\frac{1}{2} \sec^2 \mu_1)
\theta_2$.  Now we recover the harmonic edge angle from equation
\cref{eq:34}
\begin{equation}
  \label{eq:87}
  \beta_1 = \cot^{-1} \omega_k(k_1,\mu_1) = \cot^{-1}
  \frac{\omega_\varphi(\varphi_1,\mu_1)}{k_\varphi(\varphi_1,\mu_1)} = \mu_1 + 3
  \sec^2 \mu_1 \theta_2.
\end{equation}

The soliton edge simple wave problem reduces similarly to 
\begin{equation}
  \label{eq:95}
  \frac{\mathrm{d} \phit}{\mathrm{d} \oth} = \frac{1}{2} \sec^2 \mu_1
  , \quad \phit(0) = \mu_1 .
\end{equation}
Evaluating at $\oth = \theta_2$ and inserting into \cref{eq:32} gives
\begin{equation}
  \label{eq:96}
  \beta_2 \sim \mu_1 + \frac{1}{2} \sec^2 \mu_1 \theta_2 .
\end{equation}

Both \cref{eq:87} and \cref{eq:96} agree with the alternative
approach in \cite{hoefer_dark_2012}.

\subsection{Hypersonic Regime}
\label{sec:hypersonic-regime}

The hypersonic regime $M_1 \gg 1$, $\theta_2 \ll 1$ such that
$M_1 \theta_2 = \mathcal{O}(1)$, or, equivalently $0 < \mu_1 \sim
\theta_2 \ll 1$, was studied in \cite{el_two-dimensional_2009} by an
asymptotic reduction of the (2+0)D NLS equation to a (1+1)D NLS
equation.  One might naively try to utilize the small amplitude
results from the prior section with $\mu_1$ small.  However, this does
not provide the correct results.  We now recover the hypersonic
results through asymptotics of the Whitham modulation equations.  When
$0 < \omu \ll 1$, the DSW locus \cref{eq:77} yields the simple wave
relation
\begin{equation}
  \label{eq:36}
  \theta_2 \sim 2 (\mu_2 - \mu_1), \quad 0 < \mu_i \ll 1 .
\end{equation}
Since $\oth$ and $\omu$ are small, we observe from \cref{eq:73} that
$\varphi = \mathcal{O}(\omu)$.  The simple wave initial value problem for
the harmonic edge \cref{eq:91} asymptotically simplifies tremendously
to
\begin{equation}
  \label{eq:46}
  \frac{\mathrm{d} \varphi}{\mathrm{d} \omu} = 1, \quad \varphi(\mu_2) = 3
  \mu_2 - 2 \mu_1 .
\end{equation}
Solving equation \cref{eq:46} and evaluating at the harmonic wave
edge we obtain $\varphi_1 \equiv \varphi(\mu_1) = 2 \mu_2 - \mu_1$.
Inserting this result into equation \cref{eq:34} yields the harmonic
edge angle
\begin{equation}
  \label{eq:39}
  \beta_1 = \cot^{-1} \omega_k(k_1,\mu_1) = \cot^{-1}
  \frac{\omega_\varphi(\varphi_1,\mu_1)}{k_\varphi(\varphi_1,\mu_1)} \sim
  \frac{\mu_1^2 + 4 \mu_1 \theta_2 + 2\theta_2^2}{\theta_2 + \mu_1} .
\end{equation}
Utilizing the asymptotic substitution $\mu_1 \sim 1/M_1$, the result
\cref{eq:39} was also found in \cite{el_two-dimensional_2009}.

The soliton edge is analyzed in a similar way for $0 < \omu \ll 1$,
resulting in the simple wave ODE
\begin{equation}
  \label{eq:48}
  \frac{\mathrm{d} \phit}{\mathrm{d} \omu} = 1, \quad \phit(\mu_1) =
  \mu_1 ,
\end{equation}
with solution $\phit(\omu) = \omu$.  Then $\phit_2 \equiv \phit(\mu_2)
= \mu_2$ and equation \cref{eq:32} yield the soliton edge angle
\begin{equation}
  \label{eq:79}
  \beta_2 = \cot^{-1} \frac{\wt(\kt_2,\mu_2)}{\kt_2} = \cot^{-1}
  \frac{\qt(\phit_2,\mu_2)\cos \phit_2}{\qt(\phit_2,\mu_2)\sin
    \phit_2} \sim \mu_2 \sim \mu_1 + \frac{1}{2} \theta_2 ,
\end{equation}
which agrees with \cite{el_two-dimensional_2009}.

\subsection{Simple Oblique DSW}
\label{sec:simple-oblique-dsw}

\begin{figure}
  \centering
  \includegraphics[scale=0.3333]{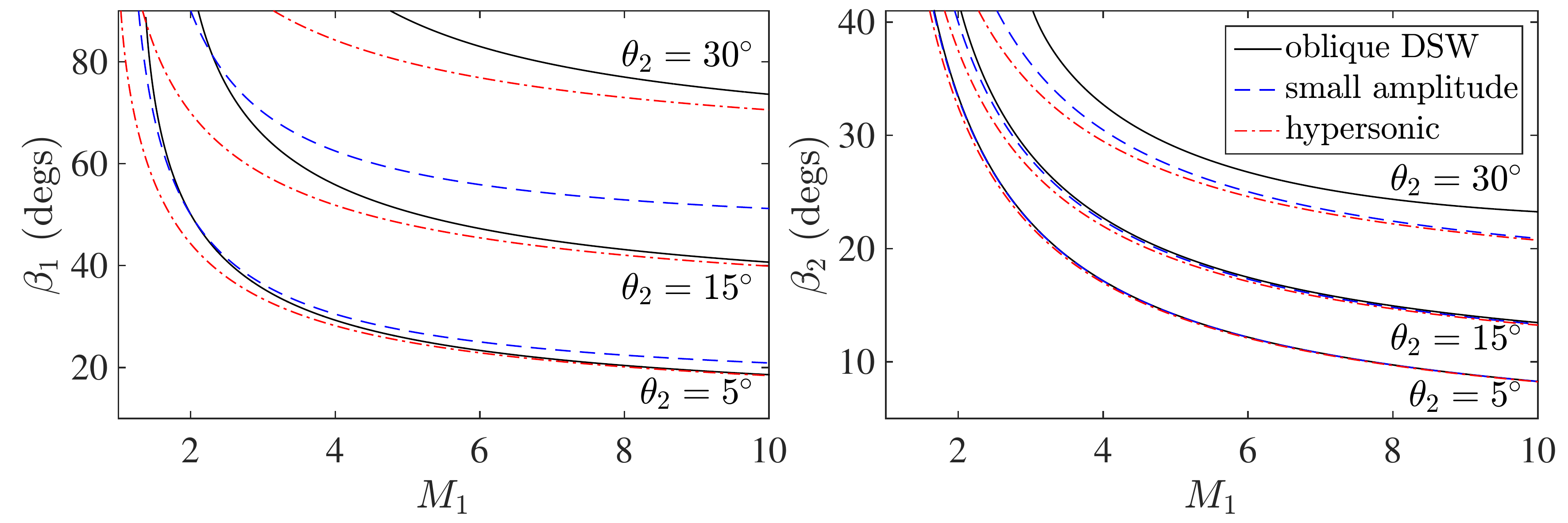}
  \caption{Oblique DSW angles versus upstream Mach number for three
    values of the flow angle $\theta_2$.  These plots highlight the
    limitations of the small amplitude (dashed) and hypersonic
    (dash-dotted) asymptotic regimes when compared with the full
    construction (solid).}
  \label{fig:asymptotics_vs_full}
\end{figure}
In order to investigate oblique DSWs across a wide parameter regime,
we numerically solve the ODEs \cref{eq:91} and \cref{eq:76}, utilizing
the DSW locus \cref{eq:77}, for the oblique DSW angles $\beta_1$ and
$\beta_2$, respectively as well as for the downstream flow properties.
The problem as formulated (cf.~\cref{fig:bvp}) involves two free
parameters, which we choose to be the upstream Mach number $M_1$ and
the downstream flow angle $\theta_2$.  These are the two natural input
parameters for supersonic flow past a corner.

\begin{figure}
  \centering
  \includegraphics[scale=0.3333]{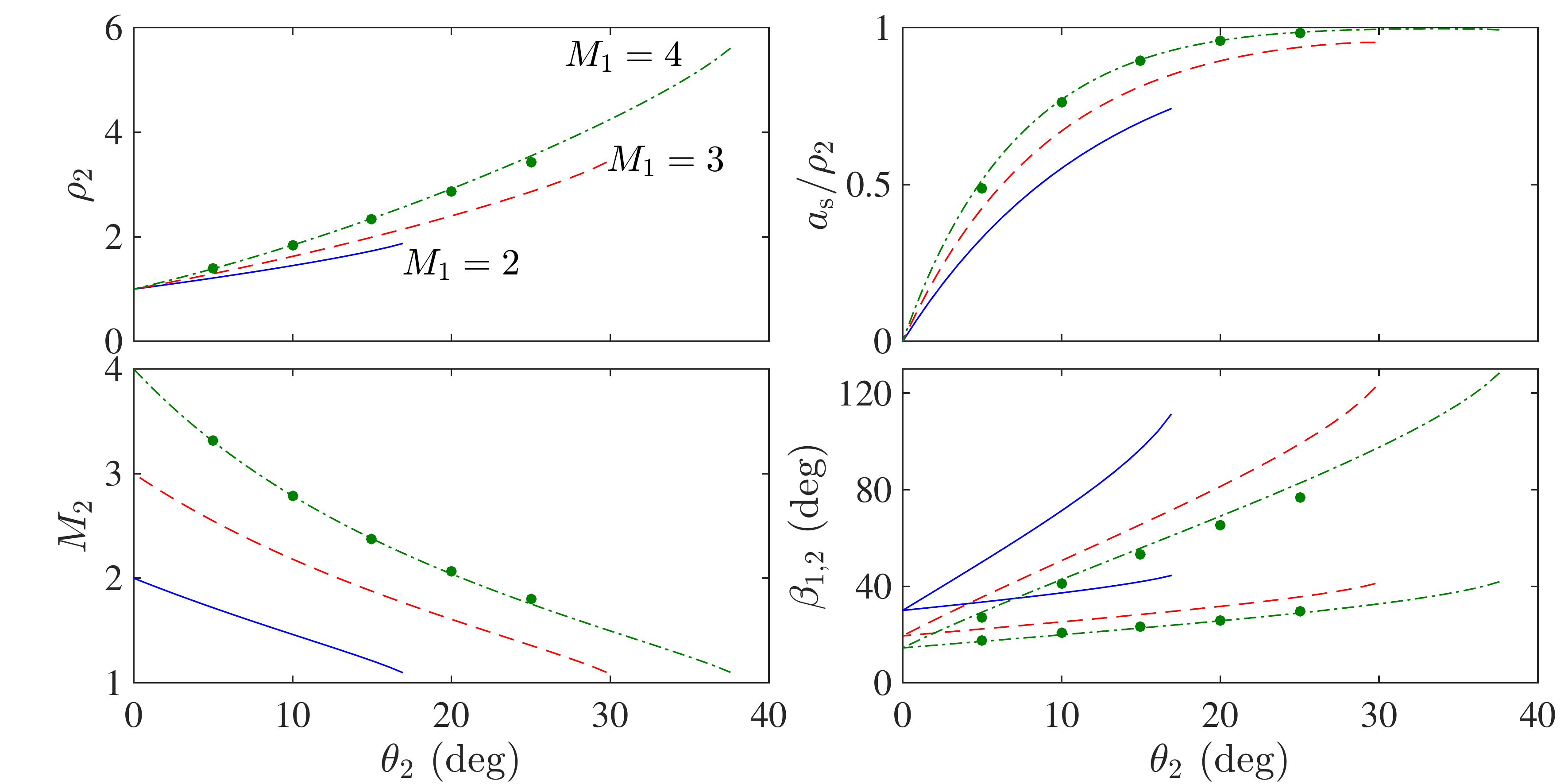}
  \caption{Oblique DSW parameters for fixed upstream Mach number $M_1
    = 2$ (solid), $M_1 = 3$ (dashed), $M_1 = 4$ (dash-dotted).  Both
    $\beta_1$ and $\beta_2$ are plotted in the lower-right panel
    ($\beta_1 > \beta_2$).  Numerically extracted quantities from
    supersonic flow over a corner are filled circles.}
  \label{fig:params_fixedM1}
\end{figure}
First we compare predictions for the oblique DSW angles $\beta_{1,2}$
with the small amplitude and hypersonic predictions from
\cref{sec:small-ampl-regime} and \cref{sec:hypersonic-regime},
respectively in \cref{fig:asymptotics_vs_full}.  Recall that both
asymptotic regimes require $0 < \theta_2 \ll 1$ and either $M_1 =
\mathcal{O}(1)$ or $M_1 \theta_2 = \mathcal{O}(1)$ in the small
amplitude or hypersonic regime, respectively.  The regimes of validity
are borne out by our computations although the soliton edge angle
$\beta_2$ exhibits better than expected agreement with the
asymptotics.

\begin{figure}
  \centering
  \includegraphics[scale=0.3333]{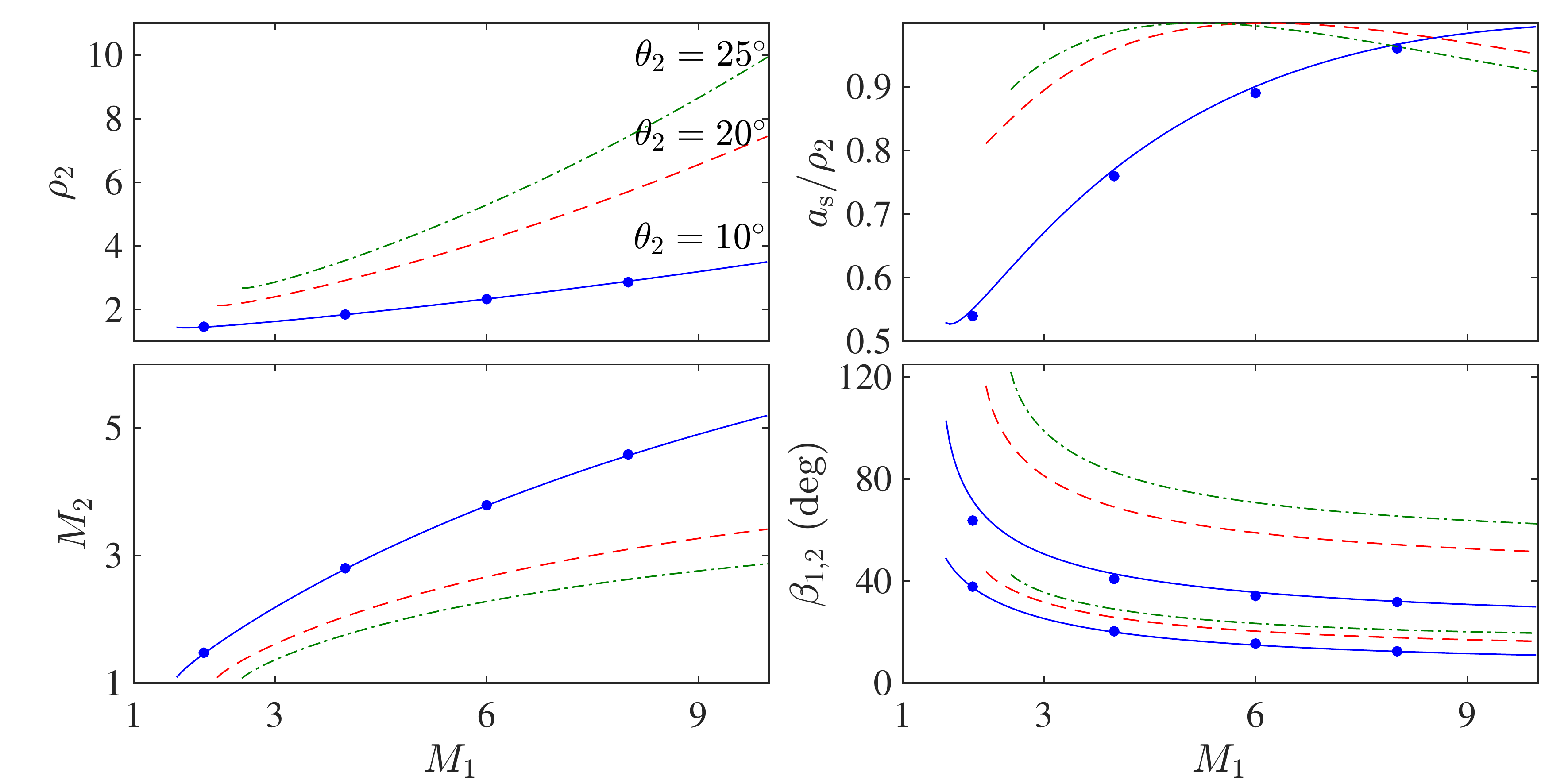}
  \caption{Oblique DSW parameters for fixed corner angle $\theta_2 =
    10^\circ$ (solid), $\theta_2 = 20^\circ$ (dashed), $\theta_2 =
    25^\circ$ (dash-dotted).  Numerically extracted quantities from
    supersonic flow over a corner are filled circles.  }
  \label{fig:params_fixedt2}
\end{figure}

Representative results for the full oblique DSW construction are shown
in \cref{fig:params_fixedM1} and \cref{fig:params_fixedt2}.  Several
features are noticeable.  In both figures, the oblique DSW ceases to
exist when the downstream flow is subsonic, $M_2 < 1$.  The
dispersionless equations \cref{eq:14} are elliptic in the subsonic
regime so our construction is no longer valid.  In
\cref{fig:params_fixedt2}, we observe that the DSW soliton edge
amplitude saturates ($a_{\rm s} = \rho_2$) for both $\theta_2 =
20^\circ$ and $\theta_2 = 25^\circ$.  This soliton amplitude
saturation corresponds to cavitation or the development of a region of
zero density.  A more careful examination shows that saturation occurs
when the downstream flow angle and DSW soliton edge angle coincide
$\theta_2 = \beta_2$.  Further increase of the flow angle $\theta_2$
requires a modification of the DSW fitting procedure.  This
modification was carried out in the hypersonic regime
\cite{el_two-dimensional_2009} but we do not do so here.

We further explore parameter space with time-dependent numerical
simulations of supersonic flow past a corner.  We incorporate a
time-dependent linear potential in the NLS equation \cref{eq:1} that
models the effect of a corner with angle $\theta_2$ moving through a
confined fluid. Initial data corresponds to the steady state within a
static confining potential.  For $t > 0$, the corner moves at a
constant speed determined by $M_1$.  A sufficiently large domain was
used to resolve the development of oblique DSWs with a grid spacing of
$0.25$ for a pseudospectral-Fourier spatial discretization and fourth
order Runge-Kutta time-stepping.  The numerical method is described in
detail in \cite{hoefer_dark_2012}.

\Cref{fig:params_fixedM1} and \cref{fig:params_fixedt2} include a
comparison of direct numerical simulations applied to the asymptotic
theory, yielding excellent agreement.  There are restrictions on the
validity of the asymptotic theory, which we now discuss in the next
subsection.

\subsection{Admissibility}
\label{sec:admissibility}

For a DSW approximated by the DSW fitting method, as herein, it must
satisfy certain causality conditions and admissibility criteria
\cite{el_dispersive_2016}.  The causality conditions for an oblique
DSW ensure that the second, dispersionless characteristic family
associated with $\lambda_+$ carries data into the DSW region.  They
are
\begin{equation}
  \label{eq:89}
  \lambda_-(0,\mu_1) < \cot \beta_1 < \lambda_+(0,\mu_1), \quad
  \lambda_+(\theta_2,\mu_2) < \cot \beta_2, \quad \beta_1 > \beta_2.
\end{equation}
These conditions have been verified in the small amplitude
\cite{hoefer_dark_2012} and hypersonic \cite{el_two-dimensional_2009}
regimes.  We have numerically verified these relations to hold for $0
< \theta_2 < \pi/2$ and $1 < M_1 < 20$ when $1 < M_2$, i.e., for
supersonic downstream flow.

The DSW fitting method breaks down when the underlying assumption of
the existence of a simple wave solution to the Whitham equations no
longer holds \cite{hoefer_shock_2014}.  Two possible mechanisms for
such a breakdown are a loss of genuine nonlinearity in the Whitham
equations or zero dispersion, which ultimately result from a loss of
monotonicity.  Because it incorporates exact reductions of the Whitham
equations at the harmonic and soliton edges, the DSW fitting method
provides a means to test for simple wave breakdown at the DSW edges.
Breakdown manifests at a DSW edge when its speed, or angle in our
case, experiences an extremum as one of the edge parameters is varied
\cite{hoefer_shock_2014,el_dispersive_2016}.  We find that
$\beta_2(M_1,M_2)$, with $M_2$ fixed, exhibits a minimum when $M_1 =
M_{\rm gnl}(M_2)$.  This minimum corresponds to the occurrence of a
linearly degenerate point or the loss of genuine nonlinearity at the
oblique DSW soliton edge.  For $M_1 > M_{\rm gnl}(M_2)$, the simple wave
assumption no longer holds, placing a restriction on the validity of
the DSW fitting method.  We have also computed $\beta_1$ with fixed
$M_2$ and $\beta_{1,2}$ with fixed $M_1$ and find no extremum.

\begin{figure}
  \centering
  \includegraphics{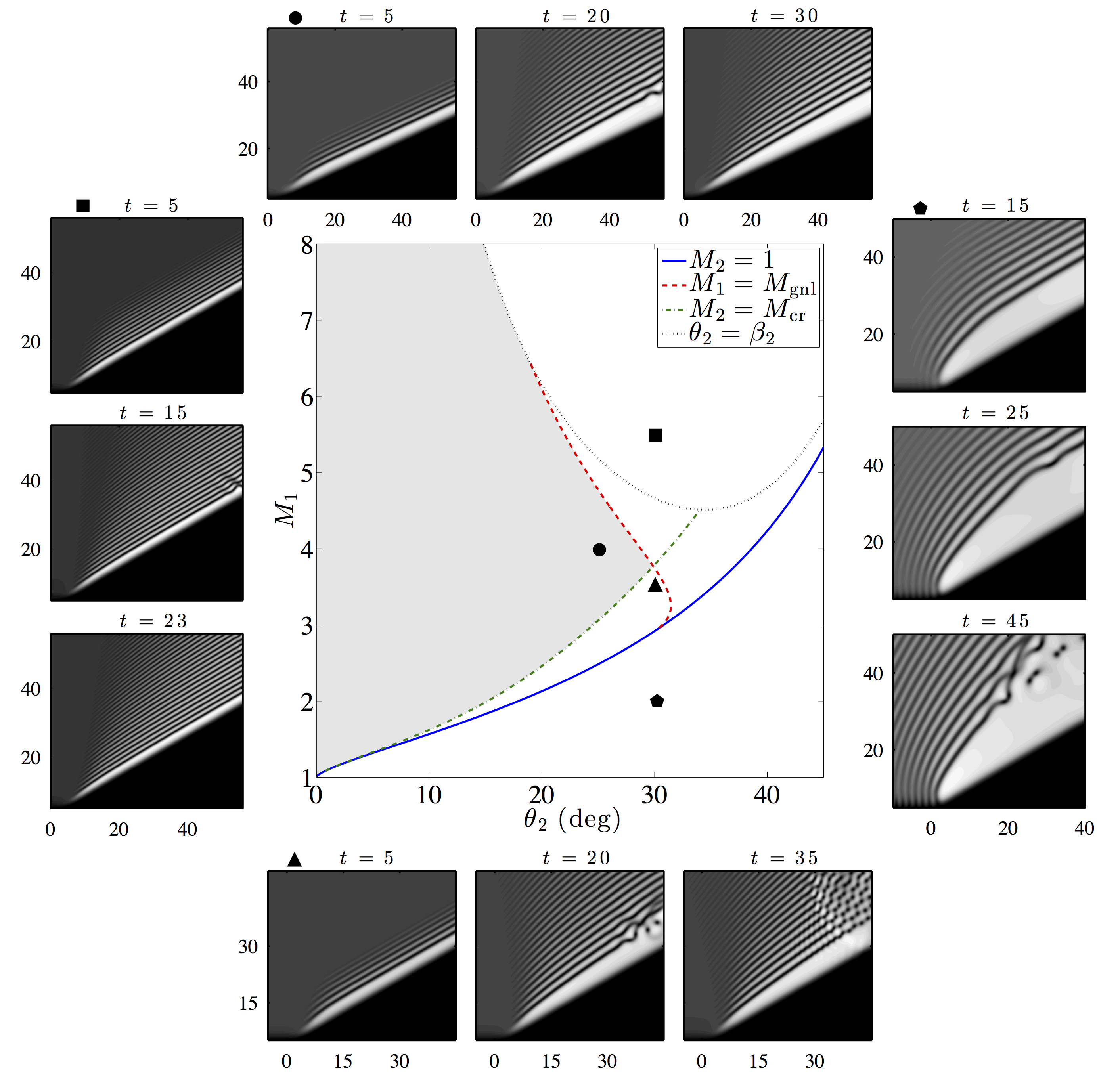}
  \caption{Phase diagram depicting the region of $(\theta_2,M_1)$
    parameter space (gray) where oblique DSWs are causal $(M_2 > 1)$,
    admissible $(M_1 < M_{\rm gnl})$, convectively unstable $(M_2 >
    M_{\rm cr})$, and non-cavitating $(\theta_2 < \beta_2)$.
    Numerical simulations of flow past corner: \mycircle--causal,
    admissible, convectively unstable, $M_1 = 4$, $\theta_2 =
    25^\circ$; \mytriangle--causal, admissible, absolutely unstable,
    $M_1 = 3.5$, $\theta_2 = 30^\circ$; \mypentagon--not causal,
    $M_1 = 2$, $\theta_2 = 30^\circ$; \mysquare--cavitation, $M_1 =
    5.5$, $\theta_2 = 30^\circ$.}
  \label{fig:phase_diagram}
\end{figure}
Although for certain parameters the oblique DSW may be causal and
admissible, it is unstable.  It is well-known that line dark solitons
are unstable to long wavelength transverse perturbations
\cite{kuznetsov_instability_1988}.  As of yet, there is not a full
analysis of DSW instability but because the oblique DSW exhibits a
dark soliton edge, the dark soliton's instability properties are
effectively inherited by the DSW.  The instability type, convective or
absolute, depends on the Mach number $M$ of the background flow
\cite{kamchatnov_stabilization_2008,hoefer_dark_2012}.  If $M > M_{\rm
  cr}$, then any localized perturbation along an oblique dark soliton,
considered at a fixed point, decays in time; i.e., the instability is
``convected away'' by the flow.  When $M < M_{\rm cr}$, absolute
instability, perturbations grow in time at any fixed point in space.
The critical Mach number $M_{\rm cr}$ depends on the normalized dark
soliton amplitude $0 < \nu = a_{\rm s}/\bar{\rho} \le 1$ and exhibits
the bounds $1 < M_{\rm cr}(\nu) \lessapprox 1.4374$ and the small
amplitude asymptotics $M_{\rm cr} = 1 + \frac{2}{9} \nu^4 +
\mathcal{O}(\nu^6)$ \cite{hoefer_dark_2012}.  For larger $\nu$,
$M_{\rm cr}(\nu)$ must be computed.  The Mach number of the oblique
DSW flow adjacent to the soliton edge is $M_2$.  Therefore, we predict
a convectively unstable oblique DSW when $M_2 > M_{\rm cr}$ and an
absolutely unstable oblique DSW otherwise.

The four conditions for a causal ($M_2 > 1$), admissible ($M_1 <
M_{\rm gnl}$), convectively unstable ($M_2 > M_{\rm cr}$), and
non-cavitating ($\theta_2 < \beta_2$) (\textit{valid}) oblique DSW,
are depicted in the phase diagram of \cref{fig:phase_diagram}.  Our
oblique DSW construction is valid across a range of upstream Mach
numbers $M_1$ and downstream flow angles $\theta_2$.  We find a
maximum valid flow angle of $\theta_2 \approx 30^\circ$ where $M_1
\approx 3.8$.

\Cref{fig:phase_diagram} shows the time evolution, in the reference
frame of the corner, of four distinct regimes of the phase diagram.
The valid region is represented by the simulation annotated with a
filled circle where an oblique DSW develops; the transverse
instability is apparent, but ultimately convects away from the corner,
leaving two distinct angles that enclose the oscillatory oblique DSW.
The simulation annotated by a filled triangle develops in a similar
fashion, but due to absolute instability, waves invade the oblique
DSW, migrating down to the corner.  A more violent instability occurs
for acausal parameter choices, represented by the filled pentagon.
The downstream flow is predicted to be subsonic and is accompanied by
the unsteady, forward propagation of dispersive waves.  Absolute
instability corresponds to the breakup of the wavetrain into vortices
that eventually overwhelm the flow.  We also include a simulation of
the cavitating regime, annotated by the filled square, where an
oblique DSW forms that is effectively ``attached'' to the corner.  In
the hypersonic regime, cavitating DSWs were shown to no longer exhibit
a dark soliton edge \cite{el_two-dimensional_2009}.  Rather, the edge
is described by a non-modulated periodic wavetrain adjacent to a
modulated region whose envelope decays to the harmonic wave edge.

While the simulations depicted in \cref{fig:phase_diagram} show the
distinction between convective and absolute instability, longer time
DSW stability can be more complicated.  In the case of convective
instability, a large number of vortices are generated and are
initially convected away from the corner.  However, these vortices can
reflect off of the boundary and interact with the DSW.  This results
in wave generation that can eventually propagate throughout the DSW,
effectively rendering it unstable for long enough evolution time.  One
way to avoid this effect is to terminate the ramp at some finite
distance away from the corner so that the generated vortices do not
reflect off the boundary.  This is what was done for the example
identified by the filled circle in \cref{fig:phase_diagram} where the
ramp was terminated at $x = 55$.

\section{Discussion and conclusion}
\label{sec:disc-concl}

We have constructed the theory of steady, oblique, two-dimensional
spatial dispersive shock waves (DSWs) in the framework of the
defocusing nonlinear Schr\"odinger (NLS) equation. This problem is of
fundamental interest as a dispersive counterpart of the classical gas
dynamics problem but also is motivated by actual physical applications
in superfluid dynamics and nonlinear optics where the NLS equation is
an accurate mathematical model.  The prototypical flow past corner
problem has been considered that elucidates the key properties of
steady oblique DSWs.  The main development of the paper is that the
theory has been constructed for general supersonic flows, without
additional simplifying assumptions (small corner angle or highly
supersonic oncoming flow) enabling one to asymptotically reduce the
description of steady oblique DSWs to solving a Riemann problem for an
integrable equation: either KdV or (1+1)D NLS
\cite{gurevich_supersonic_1995,gurevich_supersonic_1996,el_two-dimensional_2009}.
In contrast, the full (2+0)D defocusing NLS equation is not integrable
and also exhibits a number of qualitative, structural differences
compared to the above asymptotic (1+1)D models.  Thus, the previously
existing theory of steady oblique DSWs required significant
development.

In this paper, we have employed the simple wave DSW fitting method
\cite{el_resolution_2005,el_dispersive_2016}, which is based on
Whitham nonlinear modulation theory \cite{whitham_linear_1974} and is
applicable to non-integrable systems of dispersive hydrodynamics.  We
determine the definitive characteristics of a steady oblique DSW: its
locus and the bounding angles. We have identified the range of input
parameters (the Mach number of the upstream flow $M_1$ and the corner
angle $\theta_2$) for which the constructed simple-wave modulation
description is valid. The main restrictions include causality
inequalities and the condition of genuine nonlinearity of the
modulation equations.  An additional restriction of the simple wave
DSW theory specific to the defocusing (2+0)D NLS equation is the
requirement of the absence of DSW cavitation.  For sufficiently large
flow deflection angles, an additional wave structure exhibiting points
of zero density in the region adjacent to the corner can occur.

Importantly, admissible stationary oblique DSWs exhibit instability,
either convective or absolute. The convectively unstable DSWs are
effectively stable in the laboratory reference frame
\cite{kamchatnov_stabilization_2008,hoefer_theory_2009,hoefer_dark_2012}
and so can be manifested in an experiment, although a full DSW
stability theory has not yet been developed. Convective instability is
a unique property of steady oblique 2D DSWs which contrasts them to
their unsteady 1D counterparts, which are always stable. When the Mach
number of the downstream flow is smaller than a critical value
$M_{cr}$, the convective instability is replaced by absolute instability
and further, by a non-causal, subsonic regime. The boundaries of the
regions in the $\theta_2, M_1$ phase plane corresponding to the
various oblique DSW generation regimes have been identified
numerically.

The constructed theory of stationary oblique DSWs suggests several
directions of further research.  A natural next step would be the
generalization of the developed theory to more complicated geometries
such as supersonic NLS flow past an airfoil exhibiting two types of
oblique DSWs with contrasting asymptotic behaviors (see
\cite{gurevich_supersonic_1996,el_two-dimensional_2009} for the
descriptions in the frameworks of integrable asymptotic reductions).
The development of the full theory of oblique DSW stability remains an
important open problem.  Yet another closely related open problem is
the description of transonic dispersive flows.  Finally, due to the
common structure of the dispersionless equations, the analysis
performed here can be generalized to other Eulerian dispersive
hydrodynamic systems \cite{hoefer_shock_2014,el_dispersive_2016} such
as shallow water waves.

\appendix\section{Simple Wave ODEs}
\label{sec:appendix}
In this appendix, we derive the simple wave ODE \cref{eq:74}
utilizing the parametric representation \cref{eq:70,eq:71}
of the linear dispersion relation.

\subsection{Harmonic Edge}
\label{sec:harmonic-edge}

The zero amplitude modulation equations are the dispersionless
equations \cref{eq:14} for the averaged variables
$(\orho,\overline{\mathbf{u}})$ or, equivalently $(\orho,\omu,\oth)$,
coupled to the conservation of waves \cref{eq:22}, where $\omega$
corresponds to the linear dispersion relation \cref{eq:69}.  For a
simple wave, the
relations \cref{eq:75} and \cref{eq:80} hold, determining $\orho =
\orho(\omu)$ and $\oth = \oth(\omu)$.  This leaves two modulation
equations
\begin{equation}
  \label{eq:35}
  \begin{split}
    A
    \begin{bmatrix}
      \omu \\ \varphi
    \end{bmatrix}_y + &B     \begin{bmatrix}
      \omu \\ \varphi
    \end{bmatrix}_x = 0, \\
    A =
    \begin{bmatrix}
      1 & 0 \\
      q_{\omu} \sin \varphi & q_\varphi \sin \varphi + q \cos \varphi
    \end{bmatrix}, \quad
    &B =
    \begin{bmatrix}
      \lambda_+ & 0 \\
      q_{\omu} \cos \varphi & q_\varphi \cos \varphi - q \sin \varphi
    \end{bmatrix} ,
  \end{split}
\end{equation}
where we have used the parametric representation \cref{eq:70},
\cref{eq:71} with parameter $\varphi$ for the linear dispersion
relation.  This system has two characteristic speeds: $\lambda_+$, the
dispersionless characteristic, and
\begin{equation}
  \label{eq:37}
  \frac{q_\varphi \cos \varphi - q \sin \varphi}{q_\varphi \sin \varphi + q
    \cos \varphi} = \frac{\omega_\varphi}{k_\varphi} = \omega_k,
\end{equation}
the group velocity of linear waves.  A simple wave along the
$\lambda_+$ characteristic corresponds to the Prandtl-Meyer expansion
fan \cref{eq:60} in dispersionless dynamics.  For the DSW, we
seek the simple wave along the $\omega_k$ characteristic.  The goal is
to determine the appropriate wavenumber $k$ at which to evaluate the
group velocity given the Riemann initial data \cref{eq:67}.  The left
eigenvector
\begin{equation}
  \label{eq:38}
  l^T =
  \begin{bmatrix}
    q_{\omu}(\cos \varphi - \omega_k \sin \varphi) & \omega_k - \lambda_+
  \end{bmatrix},
\end{equation}
satisfies the relation $l^T B = \omega_k l^T A$.  Applying $l^T$ to
equation \cref{eq:35} results in the characteristic form
\begin{equation}
  \label{eq:40}
  l^T A
  \begin{bmatrix}
    \mathrm{d} \omu \\ \mathrm{d} \varphi
  \end{bmatrix}
  = 0 .
\end{equation}
We seek a solution in the form $\varphi = \varphi(\omu)$, yielding the
simple wave ODE
\begin{equation}
  \label{eq:41}
  \begin{split}
    \frac{\mathrm{d} \varphi}{\mathrm{d} \omu} &= \frac{q_{\omu}(\cos \varphi -
      \lambda_+ \sin \varphi)}{\sin \varphi (\lambda_+ q_\varphi + q) + \cos \varphi
      (\lambda_+ q - q_\varphi)} ,
  \end{split}
\end{equation}
that simplifies to the ODE in \cref{eq:91}.  The zero wavenumber condition at
the soliton edge $k(\varphi,\mu_2) = 0 = \omega(\varphi,\mu_2)$ corresponds
to $q(\varphi,\mu_2) = 0$ and the initial data
\begin{equation}
  \label{eq:42}
  \varphi(\mu_2) = \oth(\mu_2) + \mu_2 .
\end{equation}

\subsection{Soliton Edge}
\label{sec:soliton-edge}

For the soliton edge, we utilize the conjugate dispersion relation
\cref{eq:33}, defined parametrically according to
\begin{equation}
  \label{eq:43}
  \begin{split}
    \kt &= \qt \sin \phit, \quad \wt = \qt \cos
    \phit , \quad
    \qt = 2 \orho^{1/2} \left ( 1 - \frac{\sin^2(\phit -
        \oth)}{\sin^2 \omu} \right )^{1/2}
  \end{split}
\end{equation}
A similar calculation to that in the previous section with $q \to \qt$,
$\varphi \to \phit$ leads to the same simple wave ODE
\begin{equation}
  \label{eq:44}
  \begin{split}
    \frac{\mathrm{d} \phit}{\mathrm{d} \omu} = \frac{\qt_{\omu}(\cos \phit -
      \lambda_+ \sin \phit)}{\sin \phit (\lambda_+ \qt_{\phit} + \qt)
      + \cos \phit (\lambda_+ \qt - \qt_{\phit})} ,
  \end{split}
\end{equation}
that can be simplified to eq.~\cref{eq:94}.  The zero amplitude
condition $\kt(\phit,\mu_1) = 0 = \wt(\phit,\mu_1)$ at the harmonic
edge provides the initial condition $\qt(\phit,\mu_1) = 0$ or
\begin{equation}
  \label{eq:45}
  \phit(\mu_1) = \oth(\mu_1) + \mu_1 = \mu_1 .
\end{equation}


\begin{thebibliography}{10}

\bibitem{ablowitz_interactions_2013}
{\sc M.~J. Ablowitz and D.~E. Baldwin}, {\em Interactions and asymptotics of
  dispersive shock waves -- {Korteweg}-de {Vries} equation}, Phys. Lett. A,
  377 (2013), pp.~555--559.

\bibitem{ablowitz_dispersive_2016}
{\sc M.~J. Ablowitz, A.~Demirci, and Y.-P.~Ma}, {\em Dispersive shock
  waves in the {Kadomtsev-Petviashvili} and two dimensional
  {Benjamin-Ono} equations}, Physica D,
  333 (2016), pp.~84--98.

\bibitem{boyd_nonlinear_2013}
{\sc R.~W. Boyd}, {\em Nonlinear {Optics}}, Academic Press, Cambridge, MA,
  2013.

\bibitem{courant_supersonic_1948}
{\sc R.~Courant and K.~O. Friedrichs}, {\em Supersonic {Flow} and {Shock}
  {Waves}}, Springer-Verlag, New York, 1948.

\bibitem{dubrovin_critical_2016}
{\sc B.~Dubrovin, T.~Grave, and C.~Klein}, {\em On critical behaviour
  in generalized {Kadomtsev-Petviashvili} equations}, Physica D, 333
(2016), p.~157--170. 

\bibitem{dutton_observation_2001}
{\sc Z.~Dutton, M.~Budde, C.~Slowe, and L.~V. Hau}, {\em Observation of quantum
  shock waves created with ultra-compressed slow light pulses in a
  {Bose}-{Einstein} condensate}, Science, 293 (2001), p.~663.

\bibitem{el_resolution_2005}
{\sc G.~A. El}, {\em Resolution of a shock in hyperbolic systems modified by
  weak dispersion}, Chaos, 15 (2005), 037103.

\bibitem{el_oblique_2006}
{\sc G.~A. El, A.~Gammal, and A.~M. Kamchatnov}, {\em Oblique dark solitons in
  supersonic flow of a {Bose}-{Einstein} condensate}, Phys. Rev. Lett., 97
  (2006), 180405.

\bibitem{el_two-dimensional_2007}
{\sc G.~A. El, Y.~G. Gladush, and A.~M. Kamchatnov}, {\em Two-dimensional
  periodic waves in supersonic flow of a {Bose}-{Einstein} condensate}, J.
  Phys. A, 40 (2007), pp.~611--619.

\bibitem{el_dispersive_2016}
{\sc G.~A. El and M.~A. Hoefer}, {\em Dispersive shock waves and modulation
  theory}, Physica D, 333 (2016), pp.~11--65.

\bibitem{el_spatial_2006}
{\sc G.~A. El and A.~M. Kamchatnov}, {\em Spatial dispersive shock waves
  generated in supersonic flow of {Bose}-{Einstein} condensate past slender
  body}, Phys. Lett. A, 350 (2006), pp.~192--196.

\bibitem{el_two-dimensional_2009}
{\sc G.~A. El, A.~M. Kamchatnov, V.~V. Khodorovskii, E.~S. Annibale, and
  A.~Gammal}, {\em Two-dimensional supersonic nonlinear {S}chrodinger flow
  past an extended obstacle}, Phys. Rev. E, 80 (2009), 046317.

\bibitem{ghofraniha_shocks_2007}
{\sc N.~Ghofraniha, C.~Conti, G.~Ruocco, and S.~Trillo}, {\em Shocks in
  nonlocal media}, Phys. Rev. Lett., 99 (2007), 043903.

\bibitem{gladush_radiation_2007}
{\sc Y.~G. Gladush, G.~A.~El, A.~Gammal, and A.~M. Kamchatnov}, {\em Radiation
  of linear waves in the stationary flow of a {Bose}-{Einstein} condensate past
  an obstacle}, Phys. Rev. A, 75 (2007), 033619.

\bibitem{gurevich_supersonic_1995}
{\sc A.~V. Gurevich, A.~L. Krylov, V.~V. Khodorovskii, and G.~A. El}, {\em
  Supersonic flow past bodies in dispersive hydrodynamics}, Sov. Phys. JETP, 81
  (1995), pp.~87--96.

\bibitem{gurevich_supersonic_1996}
{\sc A.~V. Gurevich, A.~L. Krylov, V.~V. Khodorovskii, and G.~A. El}, {\em
  Supersonic flow past finite-length bodies in dispersive hydrodynamics}, Sov.
  Phys. JETP, 82 (1996), pp.~709--718.

\bibitem{gurevich_nonstationary_1974}
{\sc A.~V. Gurevich and L.~P. Pitaevskii}, {\em Nonstationary structure of a
  collisionless shock wave}, Sov. Phys. JETP, 38 (1974), pp.~291--297.
\newblock Translation from Russian of A. V. Gurevich and L. P. Pitaevskii, Zh.
  Eksp. Teor. Fiz. 65, 590-604 (1973).

\bibitem{hoefer_shock_2014}
{\sc M.~A. Hoefer}, {\em Shock waves in dispersive {Eulerian} fluids}, J. Nonl.
  Sci., 24 (2014), pp.~525--577.

\bibitem{hoefer_dispersive_2006}
{\sc M.~A. Hoefer, M.~J. Ablowitz, I.~Coddington, E.~A. Cornell, P.~Engels, and
  V.~Schweikhard}, {\em Dispersive and classical shock waves in
  {Bose}-{Einstein} condensates and gas dynamics}, Phys. Rev. A, 74 (2006),
  023623.

\bibitem{hoefer_piston_2008}
{\sc M.~A. Hoefer, M.~J. Ablowitz, and P.~Engels}, {\em Piston dispersive shock
  wave problem}, Phys. Rev. Lett., 100 (2008), 084504.

\bibitem{hoefer_theory_2009}
{\sc M.~A. Hoefer and B.~Ilan}, {\em Theory of two-dimensional oblique
  dispersive shock waves in supersonic flow of a superfluid}, Phys. Rev. A, 80
  (2009), 061601(R).

\bibitem{hoefer_dark_2012}
{\sc M.~A. Hoefer and B.~Ilan}, {\em Dark {Solitons}, {Dispersive} {Shock}
  {Waves}, and {Transverse} {Instabilities}}, SIAM Mult. Mod. Sim., 10 (2012),
  pp.~306--341.

\bibitem{jin_semiclassical_1999}
{\sc S.~Jin, C.~D. Levermore, and D.~W. McLaughlin}, {\em The semiclassical
  limit of the defocusing {NLS} hierarchy}, Comm. Pure Appl. Math., 52 (1999),
  pp.~613--654.

\bibitem{kamchatnov_flow_2010}
{\sc A.~Kamchatnov and S.~Korneev}, {\em Flow of a {Bose}-{Einstein} condensate
  in a quasi-one-dimensional channel under the action of a piston}, JETP, 110
  (2010), pp.~170--182.

\bibitem{kamchatnov_condition_2011}
{\sc A.~Kamchatnov and S.~Korneev}, {\em Condition for convective instability
  of dark solitons}, Phys. Lett. A, 375 (2011), pp.~2577--2580.

\bibitem{kamchatnov_stabilization_2008}
{\sc A.~M. Kamchatnov and L.~P. Pitaevskii}, {\em Stabilization of {solitons}
  {generated} by a {supersonic} {flow} of {Bose}-{Einstein} {condensate} {past}
  an {obstacle}}, Phys. Rev. Lett., 100 (2008), 160402.

\bibitem{kartashov_two-dimensional_2013}
{\sc Y.~V. Kartashov and A.~M. Kamchatnov}, {\em Two-dimensional dispersive
  shock waves in dissipative optical media}, Opt. Lett., 38 (2013), p.~790.

\bibitem{klein_numerical_2013}
{\sc C.~Klein and K.~Roidot}, {\em Numerical study
  of shock formation in the dispersionless {Kadomtsev-Petviashvili}
  equation and dispersive regularizations},
Physica D, 265 (2013), pp.~1--25.

\bibitem{klein_numerical_2007}
{\sc C.~Klein, C.~Sparber, and P.~A. Markowich}, {\em Numerical study
  of oscillatory regimes in the {Kadomtsev-Petviashvili} equation},
J. Nonlin. Sci., 17 (2007), pp.~429--470.

\bibitem{kuznetsov_instability_1988}
{\sc E.~A. Kuznetsov and S.~K. Turitsyn}, {\em Instability and collapse of
  solitons in media with a defocusing nonlinearity}, Sov. Phys. JETP, 67
  (1988), pp.~1583--1588.

\bibitem{lax_hyperbolic_1973}
{\sc P.~D. Lax}, {\em Hyperbolic systems of conservation laws and the
  mathematical theory of shock waves}, SIAM, Philadelphia, 1973.

\bibitem{lax_small_1983}
{\sc P.~D. Lax and C.~D. Levermore}, {\em The small dispersion limit of the
  {Korteweg}-de {Vries} equation: 1-3}, Comm. Pure Appl. Math., 36 (1983),
  pp.~253--290; 571--593; 809--830.

\bibitem{pitaevskii_bose-einstein_2003}
{\sc L.~P. Pitaevskii and S.~Stringari}, {\em Bose-{Einstein} condensation},
  Clarendon, Oxford, 2003.

\bibitem{simula_observations_2005}
{\sc T.~P. Simula, P.~Engels, I.~Coddington, V.~Schweikhard, E.~A. Cornell, and
  R.~J. Ballagh}, {\em Observations on sound propagation in rapidly rotating
  {Bose}-{Einstein} condensates}, Phys. Rev. Lett., 94 (2005), 080404.

\bibitem{venakides_long_1986}
{\sc S.~Venakides}, {\em Long time asymptotics of the {Korteweg}-de {Vries}
  equation}, Trans. Amer. Math. Soc, 293 (1986), pp.~411--419.

\bibitem{wan_dispersive_2007}
{\sc W.~Wan, S.~Jia, and J.~W. Fleischer}, {\em Dispersive superfluid-like
  shock waves in nonlinear optics}, Nat. Phys., 3 (2007), pp.~46--51.

\bibitem{whitham_non-linear_1965}
{\sc G.~B. Whitham}, {\em Non-linear dispersive waves}, Proc. Roy. Soc. Ser. A,
  283 (1965), pp.~238--261.

\bibitem{whitham_linear_1974}
{\sc G.~B. Whitham}, {\em Linear and Nonlinear Waves}, Wiley, New York, 1974.

\end{thebibliography}
\bibliographystyle{siamplain}

\end{document}